\documentclass[showpacs,prl,superscriptaddress,nofootinbib, twocolumn]{revtex4}
\usepackage{graphicx}
\usepackage{color}
\usepackage{amssymb,amsfonts,amsmath}
\usepackage{natbib}
\usepackage{epstopdf}
\usepackage{subfigure}
\usepackage{mathtools}

\DeclarePairedDelimiter{\abs}{\lvert}{\rvert}
\DeclarePairedDelimiter{\norma}{\lVert}{\rVert}
\newcommand{\sgn}{\mathop{\mathrm{sgn}}}

\begin{document}

\title{Non normal amplification of stochastic quasi-cycles.}

\author{Sara Nicoletti}
\affiliation{Universit\`{a} degli Studi di Firenze, Dipartimento di Fisica e Astronomia,
CSDC and INFN, via G. Sansone 1, 50019 Sesto Fiorentino, Italy}
\affiliation{Dipartimento di Ingegneria dell'Informazione, Universit\`{a} di Firenze,
Via S. Marta 3, 50139 Florence, Italy}
\author{Niccol\`{o} Zagli}
\affiliation{Universit\`{a} degli Studi di Firenze, Dipartimento di Fisica e Astronomia,
via G. Sansone 1, 50019 Sesto Fiorentino, Italy}
\author{Duccio Fanelli}
\affiliation{Universit\`{a} degli Studi di Firenze, Dipartimento di Fisica e Astronomia,
CSDC and INFN Sezione di Firenze, via G. Sansone 1, 50019 Sesto Fiorentino,
Italy}
\author{Roberto Livi}
\affiliation{Universit\`{a} degli Studi di Firenze, Dipartimento di Fisica e Astronomia,
CSDC and INFN Sezione di Firenze, via G. Sansone 1, 50019 Sesto Fiorentino, Italy}
\author{Timoteo Carletti}
\affiliation{naXys, Namur Institute for Complex Systems, University of Namur, Belgium}
\author{Giacomo Innocenti}
\affiliation{Dipartimento di Ingegneria dell'Informazione, Universit\`{a} di Firenze,
Via S. Marta 3, 50139 Florence, Italy}

\begin{abstract}
Stochastic quasi-cycles for a two species model of the excitatory-inhibitory type, arranged on a triangular loop, are studied. By increasing the strength of the inter-nodes coupling, one moves the system towards the Hopf bifurcation and the amplitude of the stochastic oscillations are consequently magnified. When the system is instead constrained to evolve on specific manifolds, selected so as to return a constant rate of deterministic damping for the perturbations, the observed amplification correlates with the degree of non normal reactivity, here quantified by the numerical abscissa. The thermodynamics of the reactive loop is also investigated and the degree of inherent reactivity shown to facilitate the out-of-equilibrium exploration of the available phase space.  
\end{abstract}

\pacs{02.50.Ey,05.40.-a, 87.18.Sn, 87.18.Tt, Ey,87.23.Cc, 05.40.-a} 
\maketitle

\section{Introduction}

Deterministic models are customarily invoked to reproduce in silico the intertwined dynamics of large populations of microscopic actors \cite{murray}. Stationary attractors can be identified and their inherent stability assessed, via standard techniques. By tuning an apt control parameter, a stable fixed point can turn unstable via e.g. a Hopf bifurcation, the canonical route to time periodic solutions.  The dynamical system under inspection loses stability, as a pair of complex conjugate eigenvalues - stemming from the linearized version of the problem - crosses the complex plane imaginary axis. Small-amplitude limit cycle branches from the fixed point, a dynamical transition which is intimately 
bound, under the deterministic paradigm, to positive (real parts of the) Jacobian eigenvalues. Stochastic perturbation can however play a role of paramount importance \cite{gardiner,vankampen}. Finite size corrections, arising from the system graininess, manifest as an endogenous source of disturbance, termed demographic noise. Under specific condition, the noisy contribution can shake the system from the inside yielding almost regular oscillations, the quasi-cycles, also when the 
underlying deterministic dynamics displays an asymptotically stable equilibrium, hence negative defined eigenvalues of the Jacobian matrix \cite{bartlett,mckanenewman,dauxois,bressloff,ButlerPNAS}. Quasi-cycles are often modest in size, their amplitude being set by the strength of the imposed noise source.  This fact constitutes a practical limitation, which needs to be attentively pondered, when targeting real life applications. To circumvent this impediment, we showed, in a recent work  \cite{zagli},  that giant stochastic oscillations, with tunable frequencies, can be obtained, by replicating a minimal model for quasi-cycle amplification along a directed chain. Endogenous noise fuels a coherent amplification across the array by instigating robust correlations among adjacent interacting populations.  It was argued that the observed phenomenon, explained in \cite{zagli} by resorting to the linear noise approximation,  reflected the non normal character of the imposed interaction scheme.

A linear system, in arbitrary dimension, is non normal when its governing matrix does not commute with its conjugate transpose \cite{trefethen}. Non normal systems may display a short time growth for the norm of the system state once a perturbation is  injected, even when this latter is destined to fade away at equilibrium \cite{Neubert1,Neubert2,asllani1,asllani2}. The elemental ability of a non normal system to prompt an initial rise of the associated norm, stimulated by an enduring stochastic drive, could eventually secure the sought amplification process \cite{biancalani, burioni, Hennequin}. The aim of this paper is to challenge this interpretative picture,   by considering a variant of the model presented in \cite{zagli}. More specifically, we will inspect the  dynamics of excitatory and inhibitory populations, organized in a loop, with varying coupling strength and degree of asymmetry. By forcing the system to evolve in a region of parameters where the homogeneous fixed point is stable, while freezing the (negative real part of the) largest eigenvalue to a constant amount, one can drive a sensible increase in the amplitude of the stochastic quasi-cycles by acting on the clout of non-normality. It is consequently speculated that triangular loops of the type here analyzed might define the minimal modules for self-sustained stochastic amplification in nature. Feedforwad networks with triangular architecture are often assumed in neuroscience as fundamental storage and computational units \cite{goldman, sole2}.  In this respect, our conclusions point at the crucial role that might be exerted by the non-normal nature of neuronal connectivity in the functional dynamics of cortical networks, in agreement with \cite{Hennequin}. The system being examined works as a veritable out-equilibrium thermal device under stationary conditions. The asymptotic entropy associated to steady operation increases with non normality, hinting to a novel ingredient to be included in the microscopic foundation of out-of-equilibrium thermodynamics. 

The paper is organized as follows: in the next section we will introduce the stochastic model to be probed. We will then turn to discussing its deterministic limit and study the stability of the homogeneous fixed point in the relevant parameters plane. We will also characterize the degree of non normal reactivity of the model, as witnessed by the numerical abscissa. The stochastic contribution is  then  analyzed, in Section III, under the linear noise approximation: the amplitude of the quasi-cycle will be  quantified and shown to positively correlate with the degree of reactivity displayed by the system. In Section IV, a thermodynamic interpretation is built and the concept of non normal reactivity discussed with reference to this generalized framework. 

 \section{Stochastic model}

\begin{figure}[ht!]
\includegraphics[scale=0.35]{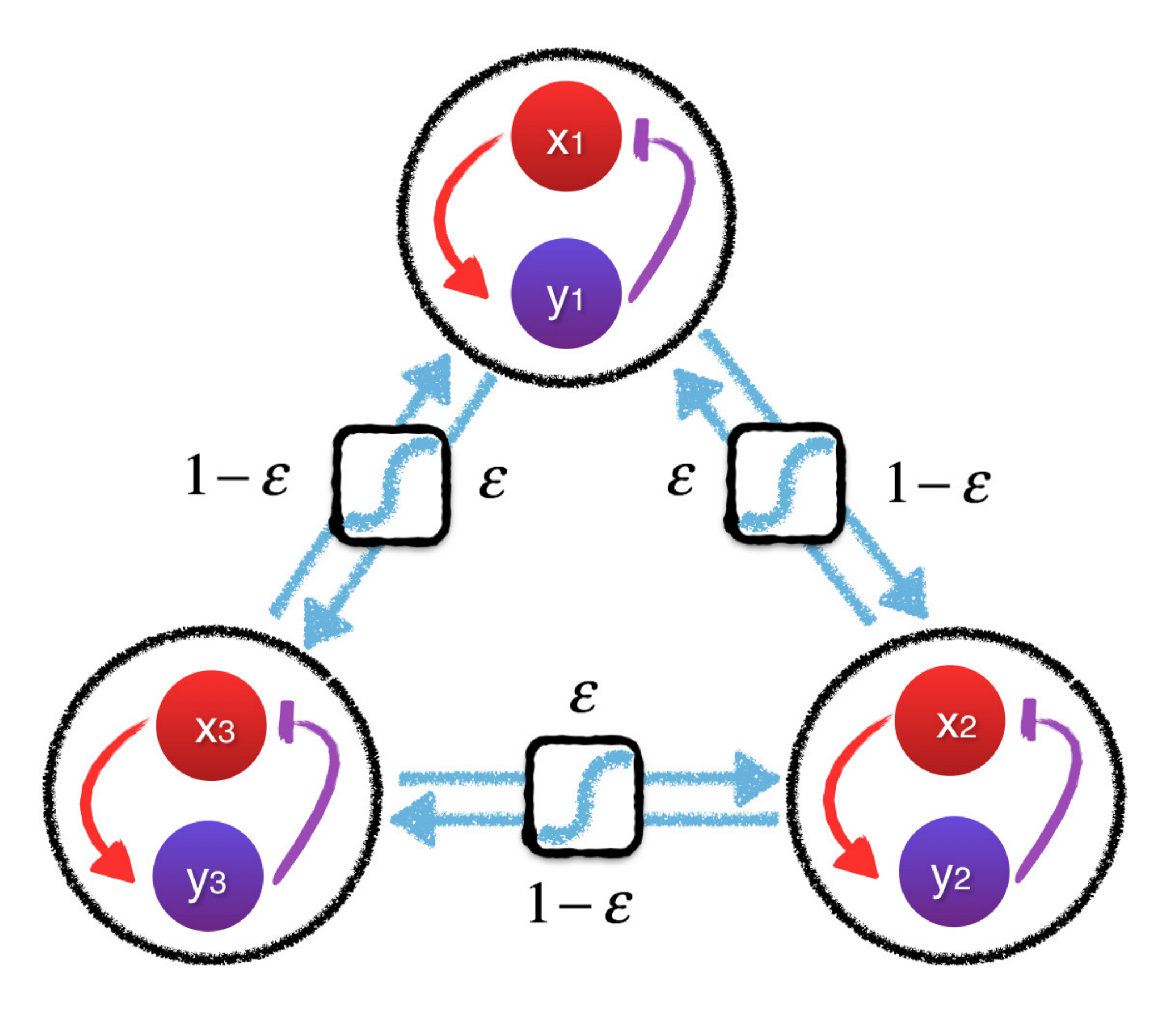}
\caption{The scheme of the model is illustrated. Two populations, labeled respectively $X$ and $Y$, are distributed on three distinct nodes of a triangular loop and therein interact via an activator-inhibitor cycle.  The nodes of the collection are coupled together, through a non linear sigmoidal function. $D$ controls the strength of the inter-nodes interaction, while $\epsilon \in [1/2 ,1]$ sets the 
degree of the coupling asymmetry. }
\label{fig1_scheme}
\end{figure}

Consider the scheme depicted in Figure \ref{fig1_scheme}.  Two populations of agents are made to mutually interact via a non linear excitatory and inhibitory circuit \cite{Zankoc}, reminiscent of the celebrated Wilson Cowan model for neuronal dynamics \cite{Cowan72,Negahbani,Cowan2016,Wallace2011}. The agents are dislocated on three different patches (nodes) defining the edges of triangular loop. The coupling among adjacent nodes is controlled by two parameters: $D$ sets the strength of the interaction, while $\epsilon \in [1/2, 1]$ stands  for the degree of imposed asymmetry. The model is  
formulated as a simple birth and death process, as we shall detail in the following. As such, it accounts for demographic stochasticity, an inevitable source of disturbance which originates from the granularity of the inspected medium. 

Denote by $X_i$ (resp. $Y_i$), one  individual of the excitatory (resp. inhibitory) species, on node $i$ ($i\in \{1,\cdots, \Omega=3\}$). Label with $n_{x_i}$ and $n_{y_i}$ the number of active excitatory and inhibitory neurons on node $i$, respectively. Furthermore, assume $V_i$ to identify the volume of the $i-$th node.
Then, the stochastic model is fully specified by the following chemical equations: 

\begin{equation}
\label{eq:equazioni chimiche vari nodi}
\begin{split}
&X_i \xrightarrow{1}  \emptyset \\
&Y_i \xrightarrow{1}  \emptyset \\
&\emptyset \xrightarrow{f(s_{x_i})} X_i \\
&\emptyset \xrightarrow{f(s_{y_i})} Y_i \\
\end{split}
\end{equation}
where $f(\cdot)=\frac{1}{1+e^{-(\cdot)}}$ is a sigmoidal function which mimics the process of neuronal activation. Networks of excitatory and inhibitory neurons represent in fact the primary computational units in the brain cortex.   Notably, inhibitory and excitatory loops, triggered by self-regulated threshold activation, are also found in genetic and metabolic cycles. Irrespectively of the specific domain of pertinence, and in light of its inherent simplicity, the 
above stochastic framework can be readily adapted to all those settings where inhibition-excitation reaction schemes are at play.

The arguments of the sigmoid function read:
\begin{equation}\label{eq:argomento sigmoide X}
s_{x_i}=-r \left(\frac{n_{y_i}}{V_i}-\frac{1}{2} \right)+D\sum_{j=1}^{\Omega} \Gamma_{ij}\left(\frac{n_{x_j}}{V_j}-\frac{n_{y_j}}{V_j}\right)
\end{equation}
\begin{equation}\label{eq:argomento sigmoide Y}
s_{y_i}=+r \left(\frac{n_{x_i}}{V_i}-\frac{1}{2} \right)+D\sum_{j=1}^{\Omega} \Gamma_{ij}\left(\frac{n_{x_j}}{V_j}-\frac{n_{y_j}}{V_j}\right)
\end{equation}
where $\Gamma_{ij}$ are the entries of the Laplacian matrix and $r$ is a local control parameter. The spatial arrangement epitomized in Figure \ref{fig1_scheme} yields the following adjacency matrix $\mathcal{A}$
\[
\mathcal{A}=
\begin{bmatrix}
0 & \epsilon & 1-\epsilon \\
1-\epsilon & 0 & \epsilon \\
\epsilon & 1-\epsilon & 0
\end{bmatrix}
\]
Then one can readily write $\Gamma_{ij}=\mathcal{A}_{ij}-k_i^{(in)}\delta_{ij}$, where $k_i^{(in)}=\sum_l \mathcal{A}_{il}$ denotes the strength (hereafter also referred to as to connectivity) of node $i$. In extended form
\begin{equation}
\label{gamma_matrix}
\Gamma=
\begin{bmatrix}
-1 & \epsilon & 1-\epsilon \\
1-\epsilon & -1 & \epsilon \\
\epsilon & 1-\epsilon & -1
\end{bmatrix}.
\end{equation}

The state of the system is completely described by the vector $\mathbf{n}=(n_{x_1},n_{y_1},...,n_{x_\Omega},n_{y_\Omega})$. 
Label with $P(\mathbf{n},t)$ the probability for the system to be in the state $\mathbf{n}$ at time $t$.
Under the Markov hypothesis, the chemical equations (\ref{eq:equazioni chimiche vari nodi})
are equivalent to a master equation for $P(\mathbf{n},t)$:
\begin{equation}\label{equazione maestra 1 nodo}
\frac{\partial P}{\partial t}(\mathbf{n},t)=\sum_{\mathbf{n}'\neq \mathbf{n}} T(\mathbf{n}|\mathbf{n}')P(\mathbf{n}',t)-T(\mathbf{n}'|\mathbf{n})P(\mathbf{n},t).
\end{equation} 

The non vanishing transition rates $T(\mathbf{n}'|\mathbf{n})$ from state $\mathbf{n}$ to state $\mathbf{n}'$, compatible with the former, are (let us observe that for a sake of clarity we only mentioned the changed variable in the new state)
\begin{equation}\label{eq:tasso di morte X vari nodi}
T(n_{x_i}-1|\mathbf{n})=\frac{n_{x_i}}{V_i}
\end{equation}
\begin{equation}\label{eq:tasso di morte Y vari nodi}
T(n_{y_i}-1|\mathbf{n})=\frac{n_{y_i}}{V_i}
\end{equation}
and 
\begin{equation}\label{eq:tasso di nascita X vari nodi}
T(n_{x_i}+1|\mathbf{n})=f\left( s_{x_i}\right)
\end{equation}
\begin{equation}\label{eq:tasso di nascita Y vari nodi}
T(n_{y_i}+1|\mathbf{n})=f\left( s_{y_i}\right).
\end{equation}
 
To proceed with the analysis we assume  $V_1$ to be large and $\gamma_i=\frac{V_i}{V1}=\mathcal{O}(1)$  $\forall i$, and seek for an approximate form of the master equation via a standard 
Kramers Moyal expansion \cite{gardiner}. The ensuing calculations are analogous to those reported in \cite{zagli} and for this reason omitted in the following. To illustrate the result of the analysis we define the macroscopic time $\tau=\frac{t}{V_1}$ and introduce the vector

 \begin{equation}\label{eq: definizione var dinamiche non lin}
 \mathbf{z}=(x_1,y_1,\dots,x_\Omega,y_\Omega)
 \end{equation}

 where $x_i=\frac{n_{x_i}}{V_i}$, $y_i=\frac{n_{y_i}}{V_i}$ are the concentrations of the active excitatory and inhibitory neurons at node $i$, with $i=1,2,3$. Notice that in our approach $V_i$ is an unspecified macroscopic parameter fixing the volume of node $i$, and, accordingly, the amplitude of the fluctuations due to demographic noise (see Eqs. (\ref{eq:eq Langevin non lineare vari nodi X}) and (\ref{eq:eq Langevin non lineare vari nodi Y})). Then, the master equation can be approximated by a Fokker-Planck equation 
\begin{equation}\label{eq: eq FP vari nodi non lineare}
\frac{\partial P}{\partial \tau}= -\sum_{i=1}^{2\Omega} \frac{\partial}{\partial z_i} A_i P +\sum_{i=1}^{2\Omega}\frac{1}{2V_1}\frac{\partial^2}{\partial z_i^2} B_{i}P
\end{equation}
with
\begin{equation}\label{eq: forze non lineari}
    \mathbf{A}=
\begin{pmatrix}
... \\
...  \\
\frac{1}{\gamma_i}\left(T\left(n_{x_i}+1|\mathbf{n}\right)-T\left(n_{x_i}-1|\mathbf{n}\right) \right) \\
\frac{1}{\gamma_i}\left(T\left(n_{y_i}+1|\mathbf{n}\right)-T\left(n_{y_i}-1|\mathbf{n}\right)  \right) \\
... \\
... \\
\end{pmatrix}
\end{equation}
and
\begin{equation}\label{eq: coupling thermal bath non lin}
    \mathbf{B}=
\begin{pmatrix}
... \\
... \\
 \frac{1}{\gamma_i^2}\left(T	\left(n_{x_i}+1|\mathbf{n}\right)+T\left(n_{x_i}-1|\mathbf{n}\right)  \right)   \\
 \frac{1}{\gamma_i^2}\left(T\left(n_{y_i}+1|\mathbf{n}\right)+T\left(n_{y_i}-1|\mathbf{n}\right)  \right)  \\
... \\
... \\ 
\end{pmatrix}.
\end{equation}
The Fokker-Planck equation \eqref{eq: eq FP vari nodi non lineare} is equivalent to the following nonlinear Langevin equations for the stochastic concentrations of the involved species 
\begin{equation}\label{eq:eq Langevin non lineare vari nodi X}
\frac{d}{d\tau}x_i=\frac{1}{\gamma_i}\left[f\left(s_{x_i}\right)-x_i \right]+\frac{1}{\sqrt{V_1}}\frac{1}{\gamma_i}\sqrt{x_i+f\left(s_{x_i}\right)}\lambda_i^{(1)}
\end{equation}
\begin{equation}\label{eq:eq Langevin non lineare vari nodi Y}
\frac{d}{d\tau}y_i=\frac{1}{\gamma_i}\left[f\left(s_{y_i})-y_i\right)  \right]+\frac{1}{\sqrt{V_1}}\frac{1}{\gamma_i}\sqrt{y_i+f\left(s_{y_i}\right)}\lambda_i^{(2)}
\end{equation}
where $<\lambda_i^{(l)}\left(\tau\right)>=0$ and $<\lambda_i^{(l)}\left(\tau\right)\lambda_j^{(m)}\left(\tau\right)>=\delta_{ij}\delta_{lm}\delta\left(\tau-\tau'\right)$ with $i,j=1,\dots,\Omega$ and $l,m=1,2$.

\section{Deterministic limit}

In the limit $V_1 \rightarrow +\infty$ one readily obtains the following deterministic equations
\begin{align}
\label{nl_sistem1}
\dot{x}_i &= \frac{1}{\gamma_i}\left[f(s_{x_i})-x_i\right] \\
\label{nl_sistem2}
\dot{y}_i &= \frac{1}{\gamma_i}[f(s_{y_i})-y_i]
\end{align}
where the dot stands for the derivative with respect to the macroscopic time $\tau$. Equations (\ref{nl_sistem1})-(\ref{nl_sistem2}) are complemented by the self-consistent conditions:
\begin{align}
s_{x_i}&=-r\biggl(y_i-\frac{1}{2}\biggr)+D\sum_{j=1}^{\Omega}\Gamma_{ij}(x_j-y_j) \\
s_{y_i}&=r\biggl(x_i-\frac{1}{2}\biggr)+D\sum_{j=1}^{\Omega}\Gamma_{ij}(x_j-y_j).
\end{align}

System (\ref{nl_sistem1})-(\ref{nl_sistem2}) admits a homogeneous fixed point $x_i=y_i=\frac{1}{2}$, $\forall i$. To assess its stability, we proceed by linearizing the dynamics around the aforementioned equilibrium. To this end we set $x_i=\frac{1}{2}+\delta x_i$, $y_i=\frac{1}{2}+\delta y_i$ and expand in power of the perturbation amounts. By arresting the expansion to the first order, one obtains the 
following system of linear equations:
\[
\begin{cases}
\delta \dot{x_i}=\frac{1}{\gamma_i}\biggl[-\delta x_i-\frac{r}{4}\delta y_i+\frac{D}{4}\sum_{j=1}^\Omega\Gamma_{ij}(\delta x_j-\delta y_j)\biggr] \\
\delta \dot{y_i}=\frac{1}{\gamma_i}\biggl[-\delta y_i+\frac{r}{4}\delta x_i+\frac{D}{4}\sum_{j=1}^\Omega\Gamma_{ij}(\delta x_j-\delta y_j)\biggr]
\end{cases}
\]
which can be cast in matricial form as
\begin{equation}\label{eq: pert lineari deterministiche}
    \frac{d}{d\tau}\delta\mathbf{x}=\mathcal{J}\delta\mathbf{x}
\end{equation}
where 
\begin{equation}
\label{J}
\mathcal{J}=
\begin{bmatrix}
L_1 & M_1 & N_1 \\
N_2 & L_2 & M_2 \\
M_3 & N_3 & L_3 
\end{bmatrix}
\end{equation}
\begin{equation}
    L_i=\frac{1}{\gamma_i}
\begin{bmatrix}
-1-\frac{D}{4} & -\frac{r}{4}+\frac{D}{4} \\
\frac{r}{4}-\frac{D}{4} & -1+\frac{D}{4}
\end{bmatrix}
\end{equation}
\begin{equation}
    M_i=\frac{1}{\gamma_i}
    \begin{bmatrix}
    \frac{D\epsilon}{4} & -\frac{D\epsilon}{4} \\
    \frac{D\epsilon}{4} & -\frac{D\epsilon}{4}
    \end{bmatrix}    
\end{equation}
\begin{equation}
    N_i=\frac{1}{\gamma_i}
    \begin{bmatrix}
    \frac{D(1-\epsilon)}{4} & -\frac{D(1-\epsilon)}{4} \\
    \frac{D(1-\epsilon)}{4} & -\frac{D(1-\epsilon)}{4}
    \end{bmatrix}.
\end{equation}

To compute the eigenvalues of the Jacobian, and eventually elaborate on the stability of the equilibrium solution, we introduce the eigenvectors $\boldsymbol{\phi}^{(\beta)}$ of the Laplacian matrix:

 \begin{equation}
      \Gamma \boldsymbol{\phi}^{(\beta)}=\Lambda^{(\beta)}\boldsymbol{\phi}^{(\beta)}, \quad \beta=1,...,\Omega 
 \end{equation}
 where $\Lambda^{(\beta)}$ are the associated eigenvalues. We can then decompose the perturbation on the basis of the eigenvectors which corresponds to setting:

\begin{equation}\label{eq: decomposizone X}
\delta x_i=\sum_{\beta=1}^{\Omega} c_\beta \exp\biggl(\frac{\lambda_\beta}{\gamma_i} \tau\biggr) \phi_{i}^{(\beta)}
\end{equation}
\begin{equation} \label{eq: decomposizione Y}
\delta y_i=\sum_{\beta=1}^{\Omega} b_\beta \exp\biggl(\frac{\lambda_\beta}{\gamma_i} \tau\biggr) \phi_{i}^{(\beta)} 
\end{equation}
where $c_\beta,b_\beta,\lambda_\beta$ are constants and $\lambda_\beta$ sets the rate of the exponential growth (or damping), as obtained under the linear approximation. Inserting the above ansatz into the governing equation and performing the calculation, one readily gets: 

\begin{equation}
\begin{pmatrix}
\lambda_\beta+1-\frac{D}{4}\Lambda^{(\beta)} &  \frac{r}{4}+\frac{D}{4}\Lambda^{(\beta)} \\
-\frac{r}{4}-\frac{D}{4}\Lambda^{(\beta)} & \lambda_\beta+1+\frac{D}{4}\Lambda^{(\beta)}
\end{pmatrix}
\begin{pmatrix}
c_\beta \\ b_\beta
\end{pmatrix}
= 0.
 \end{equation}

A non trivial solution of the above system exists, provided the determinant of the associated matrix is identically equal to zero, or stated it differently, if the rates $\lambda_\beta$ solve the 
quadratic equation

\begin{equation}
    \lambda_\beta^2+2\lambda_\beta+1+\frac{r}{16}(r+2D\Lambda^{(\beta)})=0
\end{equation}

that yields the closed formula
\begin{equation}
\lambda_\beta=-1\pm \sqrt{-\frac{r}{16}(r+2D\Lambda^{(\beta)})}, \quad \beta=1,...,\Omega.
\end{equation}
The eigenvalues  of Laplacian $\Gamma$, specified in eq. (\ref{gamma_matrix}) reads
$\Lambda^{(1)}=0$, $\Lambda^{(2,3)}=-\frac{3}{2}\pm i\frac{\sqrt{3}}{2}(1-2\varepsilon)$.  Hence, 
it is immediate to get:
\begin{equation}
\label{lambda12}
\lambda_{1,2}=-1 \pm \frac{r}{4}i=-1 \pm i\omega_0.
\end{equation}
The second eigenvalue $ \Lambda^{(2)} $ yields
\begin{equation}
\lambda_{3,4}=-1\pm \frac{1}{2}\sqrt{z}
\end{equation}
and $ \Lambda^{(3)} $
\begin{equation}
\lambda_{5,6}=-1 \pm \frac{1}{2}\sqrt{\bar{z}}
\end{equation}
with
\begin{equation}\label{eqn:z}
z=\frac{r}{4}\biggl[3D-r+\sqrt{3}Di(1-2\epsilon)\biggr],
\end{equation}

and where $\bar{z}$ stands for the complex conjugate of $z$.

By separating the real and imaginary parts returns
\begin{equation}
\lambda_{3,4}=-1\pm \frac{1}{2}\sqrt{\frac{\abs{z}+Re{z}}{2}}\pm i\sgn(Im z)\omega_1
\end{equation}
\begin{equation}
\lambda_{5,6}=-1\pm \frac{1}{2}\sqrt{\frac{\abs{z}+Re{z}}{2}}\pm  i\sgn(Im \bar{z})\omega_1
\end{equation}
where
\begin{equation}
    \omega_1=\frac{1}{2}\sqrt{\frac{\abs{z}-Re{z}}{2}}.
\end{equation}
We also define $\alpha$ as the supremum of the real part of the spectrum of $\mathcal{J}$, in formula: 
\begin{equation}
    \alpha=\sup_\beta Re\left(\lambda_\beta \right) =-1+ \frac{1}{2}\sqrt{\frac{\abs{z}+Re{z}}{2}}.
\end{equation}
A straightforward calculation allows one to  isolate the domain in the plane ($\epsilon$, $D$) where the homogeneous fixed point proves stable. The stability is enforced by setting $D<D_c$ where:

\begin{equation}
\label{Dc}
D_c=\frac{-12+\sqrt{144+(4r^2+64)3\bigl(\epsilon-\frac{1}{2}\bigr)^2}}{\frac{3}{2}r\bigl(\epsilon-\frac{1}{2}\bigr)^2}.
\end{equation}

This is a decreasing function of $\epsilon$, suggesting that asymmetry anticipates the onset of the instability. Furthermore, $D_c$ displays a minimum in $r$, and the critical value $D_c$ diverges for $r \rightarrow 0$. It is therefore possible to select arbitrarily large values of $D$, provided $r$ is sufficiently small, while still constraining the system in the region of stable homogeneous fixed point. 


The set of computed eigenvalues exhibits two distinct imaginary contributions, for $\varepsilon \neq \frac{1}{2}$, and $D<D_c$: $\omega_0=r/4$, as introduced in equation (\ref{lambda12}), and 
$\omega_1$ associated to the remaining set of eigenvalues and defined as follows:

 \begin{equation}
    \omega_1=\frac{1}{4}\sqrt{\frac{r[(3D-r)^2+3D^2(1-2\epsilon)^2]-r(3D-r)}{2}}.
\end{equation}

Interestingly, the frequency $\omega_1$ can be both smaller or bigger than $\omega_0$: indeed it is possible to show that $\omega_1>\omega_0$ if $D>D^*=4r/(1-2\epsilon)^2$. If $D=D^*$,
$\omega_0=\omega_1$.  In the limiting condition of a symmetric loop, $\varepsilon=\frac{1}{2}$, the Laplacian displays a real spectrum. More specifically, $z=\frac{r}{4}(3D-r)$ is real. Thus, $\lambda_{3,4}=\lambda_{5,6}=-1 \pm \frac{1}{2}\sqrt{\frac{r}{4}\left( 3D-r \right) }$. In this case, $D_c=\frac{r}{3}+\frac{16}{3r}$. For, $D<\hat{D}=\frac{r}{3}$, the system is stable and two frequencies are active, $\omega_0$ and $\omega_1=\frac{1}{2}\sqrt{\abs{\frac{r}{4}(3D-r)}}$.
Conversely, for $\hat{D}<D<D_c$ the system is stable but the frequency $\omega_1$ disappears. For any choice of $\epsilon$, at  $D=D_c$, two complex conjugate eigenvalues cross the vertical imaginary axis, signaling a Hopf bifurcation and the consequent inception of a limit cycle. In the following, we shall operate in the region of the plane $(\epsilon, D)$ where the system is predicted to stably converge to a homogeneous equilibrium, obtained by replicating on each node of the collection the trivial fixed point $(1/2,1/2)$.

The fate of any imposed perturbation is eventually set by the spectrum $\sigma\left(\mathcal{J}\right)$ of the Jacobian matrix $\mathcal{J}$, the matrix that governs the linear dynamics of the system around the examined equilibrium. 
Perturbations fade away when $\alpha<0$ -- for our specific case study, this amounts to setting $D<D_c$ -- and the system converges back to its deputed equilibrium. A transient growth of the perturbation can however be seen, at short times, if $\mathcal{J}$ is {\it non normal} and {\it reactive}. A matrix is said non-normal, if it does not commute with its adjoint
\cite{trefethen}. Asymmetry,  as reflecting a non trivial balance between intrinsic dynamics and enforced non local couplings, is thus a necessary condition for non normality to emerge. Since, in our case, $\mathcal{J}$ is real, taking the adjoint is identical to considering the transpose
 of the matrix. In formulae, $\mathcal{J}$ is non-normal, provided $[\mathcal{J},\mathcal{J}^T] \equiv \mathcal{J} \mathcal{J}^T - \mathcal{J}^T \mathcal{J} \neq 0$, where the apex $T$ identifies the transpose operation. It is immediate to conclude that the matrix  $\mathcal{J}$, as defined in (\ref{J}), is non-normal when $D>0$. 
 
A straightforward manipulation \cite{trefethen, asllani1} yields the following equation for  the evolution of the norm of the perturbation 
$\norma{\delta\mathbf{x}}$:
\begin{equation}
\frac{d\norma{\delta \mathbf{x}}}{d\tau}=\frac{\delta\mathbf{x}^T\mathcal{H(J)}\delta\mathbf{x}}{\norma{\delta \mathbf{x}}}
\end{equation}
where $\mathcal{H(J)}=\frac{\mathcal{J}+\mathcal{J}^T}{2}$ stands for Hermitian part of $\mathcal{J}$. The evolution of the perturbation, at short times, is intimately related to the 
so called {\it numerical abscissa}, $w=\sup\sigma\bigl(\mathcal{H(J)}\bigr)$. If $w>0$, the system is termed {\it reactive}, and perturbation may display an initial, transient growth. In this paper, we are interested in 
shedding light on the interplay between reactivity, i.e. the inherent ability of the system to yield a short time enhancement of a deterministic perturbation, and the stochastic contribution stemming from 
demographic fluctuations. As we shall see, the amplification of  quasi-cycles,  self-sustained oscillations driven by granularity, correlates with the degree of reactive non normality, as displayed by the system in its 
linearized version. To proceed in the analysis, we set to compute the eigenvalues of $\mathcal{H(J)}$ and get the following closed expression for the reactivity index $w$:
\begin{equation}
\label{omega}
w=    -1 + \frac{D}{4}\sqrt{3(\epsilon^2-\epsilon+1)}.
\end{equation}
Hence, $w>0$ when $D>D_{NN}=\frac{4}{\sqrt{3(\epsilon^2-\epsilon+1)}}$. Notice that $D_{NN}$, the lower bound in $D$ for the onset of a reactive response, is independent of $r$, and solely function of $\epsilon$.

\begin{figure}[ht!]
\centering
{\includegraphics[scale=0.28]{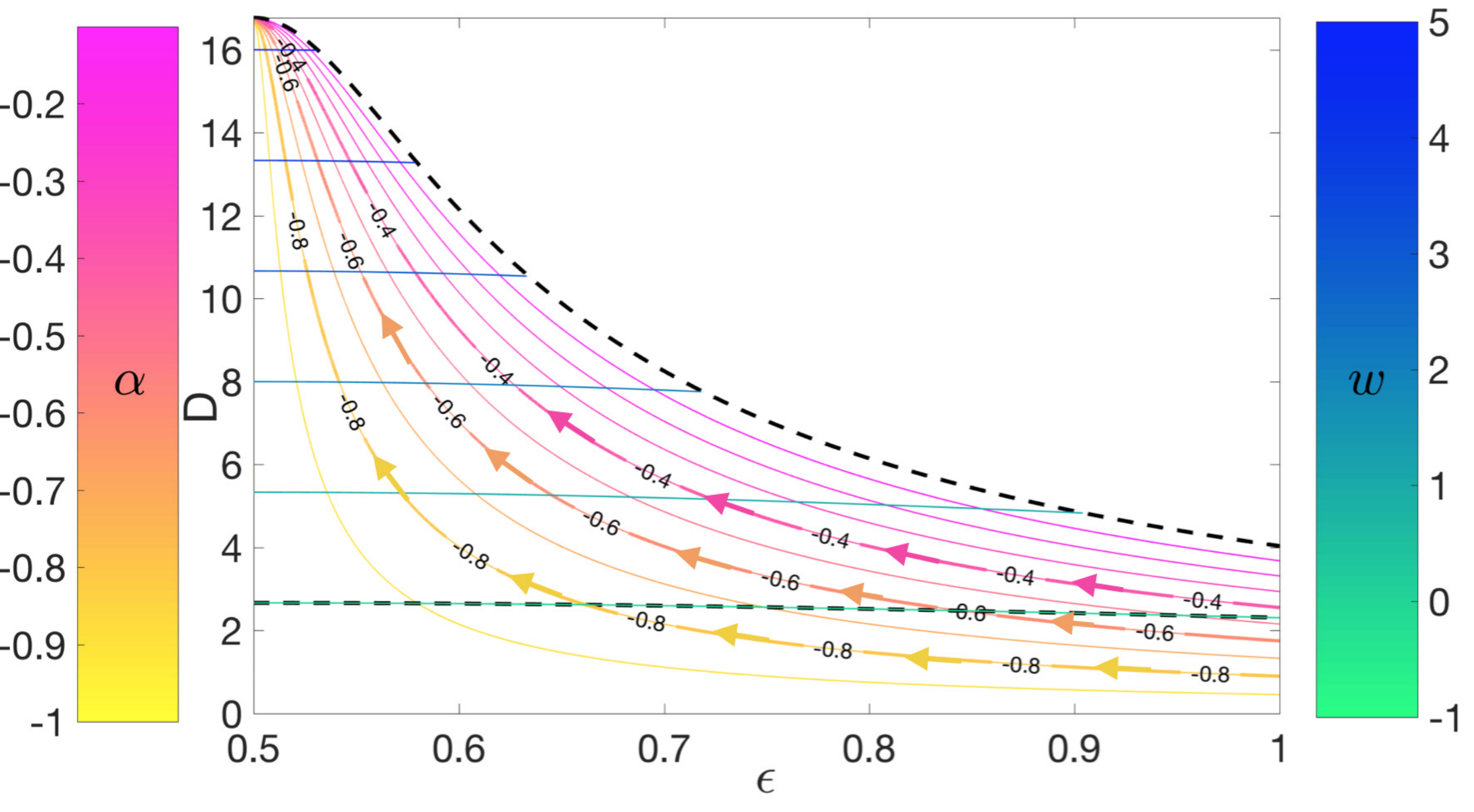}} \\
\caption{The domain of stability of the homogeneous equilibrium is depicted in the reference plane ($\epsilon$, $D$) and for $r=50$. The upper dashed curve stands for the Hopf bifurcation. Above the lower dashed curve   
the non-normal Jacobian matrix  $\mathcal{J}$ is found to be reactive.  Iso-$\alpha$ and iso-$w$ curves are traced, and colored with apt codes which reflect their associated level, as specified in the annexed bars.}
\label{fig:fig2}
\end{figure}

The above results are summarized in Figure \ref{fig:fig2}, were the boundaries of stability are depicted in the reference plane $(\epsilon,D)$, for a fixed, although representative, value of $r$. The upper dashed line stands for $D_c$
as given in equation (\ref{Dc}). The lower dashed line refers instead to $D_{NN}$ and marks the boundary of the domain where matrix  $\mathcal{J}$ is found to be reactive. Level curves traced at constant values of $\alpha$ 
(see color bar depicted on the left) and $w$ (refer to the color bar reported on the right) foliate the scanned portion of the plane. Moving along iso-$\alpha$ lines implies freezing the rate of exponential damping of the perturbation to a constant value, or, stated it differently visiting the subset of points that are, to some extent, equidistant from the frontier of the Hopf bifurcation. When crawling on  iso-$w$ lines, instead, one forces constant the (largest) rate of short time growth, as seeded by reactive non-normality.  While it is straightforward to obtain a closed analytical expression for iso-$w$ curves, upon trivial inversion of equation (\ref{omega}), more tricky proves the calculation that yields an explicit representation of iso-$\alpha$ lines. Label with $\bar{\alpha}<0$ the selected  iso-$\alpha$. Then, after a cumbersome derivation, one gets the following expression for $D$, as function of both $\bar{\alpha}$ and $\epsilon$

\begin{widetext}
\begin{equation}
D=\frac{-12(\bar{\alpha}+1)^2r+\sqrt{144(\bar{\alpha}+1)^4r^2+[64(\bar{\alpha}+1)^4+4(\bar{\alpha}+1)^2r^2]\frac{3}{4}r^2(2\epsilon-1)^2}}{\frac{3}{8}r^2(2\epsilon-1)^2}
\end{equation}
\end{widetext}
which is employed for tracing  the iso-$\alpha$ lines displayed in Figure \ref{fig:fig2}. Starting from this setting, we shall hereafter elaborate on the role of non-normality in a stochastic framework. To anticipate our findings, we will prove that the amplitude of noise driven oscillations grows with the degree of reactivity.

\section{Linear noise approximation}

To quantify the role of stochastic fluctuations around the deterministic equilibrium, we shall operate under the linear noise approximation. In concrete, we rewrite the stochastic densities $x_i$ and $y_i$, for all nodes of the collection, as the sum of two distinct contributions: the deterministic fixed point, on the one side,  and a stochastic perturbation, on the other. This latter is assumed to be modulated by a scaling factor $1/\sqrt{V_i}$, which follows the central limit theorem. 
In formulae, we postulate:

\begin{align}
x_i=\frac{1}{2}+\frac{\xi_i}{\sqrt{V_i}} \\
y_i=\frac{1}{2}+\frac{\eta_i}{\sqrt{V_i}},
\end{align}
and introduce 
\begin{equation}\label{eq: def variabili dinamiche lin}
\boldsymbol{\zeta}=(\xi_1,\eta_1,\ldots,\xi_{\Omega},\eta_{\Omega})
\end{equation}
to label the vector of fluctuations. Inserting
the above ansatz in the governing master equation and performing 
the expansion at the first order in $1/\sqrt{V_1}$ (see \cite{zagli} for details about the technical steps involved in the calculations), one eventually gets the following set of linear Langevin equations:

\begin{equation}\label{eqn:langevin}
\frac{d}{d\tau}\zeta_i=(\mathcal{J}\boldsymbol{\zeta})_i+\lambda_i
\end{equation}
where $\boldsymbol{\lambda}$ stands for a Gaussian noise that satisfies the following conditions
\begin{equation}
<\boldsymbol{\lambda}>=0,
\end{equation}
\begin{equation}
<\lambda_i(\tau)\lambda_j(\tau')>=\mathcal{B}_{ij}\delta(\tau-\tau').
\end{equation}
The diffusion matrix $\mathcal{B}_{ij}$ is defined by its diagonal elements
\begin{equation}\label{eq: def coupling lineare}
\mathcal{B}_i=\biggl(\frac{1}{\gamma_1},\frac{1}{\gamma_1},\ldots,\frac{1}{\gamma_{\Omega}},\frac{1}{\gamma_{\Omega}}\biggr).
\end{equation}
When the volumes are equal, the diffusion matrix simply reduces to $\mathcal{B}_{ij}=\delta_{ij}$.

The above Langevin equations \eqref{eqn:langevin} admits an equivalent formulation in terms of an associated Fokker-Planck equation which can be formally cast in the form:
\begin{equation}\label{eqn:fokker}
\frac{\partial}{\partial \tau}\Pi=-\sum_{i=1}^{2\Omega}\frac{\partial}{\partial\zeta_i}(\mathcal{J}\boldsymbol{\zeta})_i\Pi+\frac{1}{2}\frac{\partial^2}{\partial\zeta_i^2}\mathcal{B}_i\Pi.
\end{equation}
This latter describes the evolution of probability distribution $\Pi(\boldsymbol{\zeta},\tau)$ of the fluctuations.
\\
The solution  at any time of the above Fokker-Planck equation is a multivariate normal distribution 
\begin{equation}\label{eq: probabilità lineare}
    \Pi (\boldsymbol{\zeta},\tau)=\frac{1}{\sqrt{(2\pi)^{2\Omega}\abs{\mathcal{C}}}}\exp{\left\{ -\frac{1}{2}\left(\boldsymbol{\zeta}-<\boldsymbol{\zeta}>\right)^T\mathcal{C}^{-1}\left(\boldsymbol{\zeta}-<\boldsymbol{\zeta}>\right) \right\}  }                
\end{equation}
where $\abs{\mathcal{C}}$ is the determinant of the correlation matrix. The sought probability distribution $\Pi(\boldsymbol{\zeta},\tau)$ is hence completely characterized in terms of the first and second moments of the fluctuations, $<\zeta_i>$ and $<\zeta_l\zeta_m>$. These latter quantities obey the following differential equations \cite{zagli}:

\begin{widetext}
\begin{align}
\label{moments}
&\frac{d}{d\tau}<\zeta_i>=\left(\mathcal{J}\boldsymbol{\zeta} \right)_i  \notag \\
&\frac{d}{d\tau}<\zeta_l^2>=2<(\mathcal{J}\boldsymbol{\zeta})_l\zeta_l>+\mathcal{B}_l=2\sum_{j=1}^{2\Omega}\mathcal{J}_{lj}<\zeta_l\zeta_j>+\mathcal{B}_l \\
&\frac{d}{d\tau}<\zeta_l\zeta_m>=<(\mathcal{J}\boldsymbol{\zeta})_l\zeta_m>+<(\mathcal{J}\boldsymbol{\zeta})_m\zeta_l>=\sum_{j=1}^{2\Omega}\mathcal{J}_{lj}<\zeta_m\zeta_j>+\mathcal{J}_{mj}<\zeta_l\zeta_j>. \notag
\end{align}
\end{widetext}

The stationary moments can be analytically computed by setting to zero the time derivatives
on the left hand side of equations (\ref{moments}) and solving the system that is consequently obtained. 
The first moments are immediately found to be identically equal to zero asymptotically. Determining the second moments 
implies dealing with a linear system, which can be drastically simplified, by invoking translation invariance 
across the loop. In particular,  $<\zeta_i^2>$ take two distinct values, respectively reflecting the  typical amplitude of the  fluctuations, as displayed by excitators and inhibitors.  

\begin{figure}
 \centering
 \begin{tabular}{l}
 \textbf{a} \\
 {\includegraphics[width=9cm]{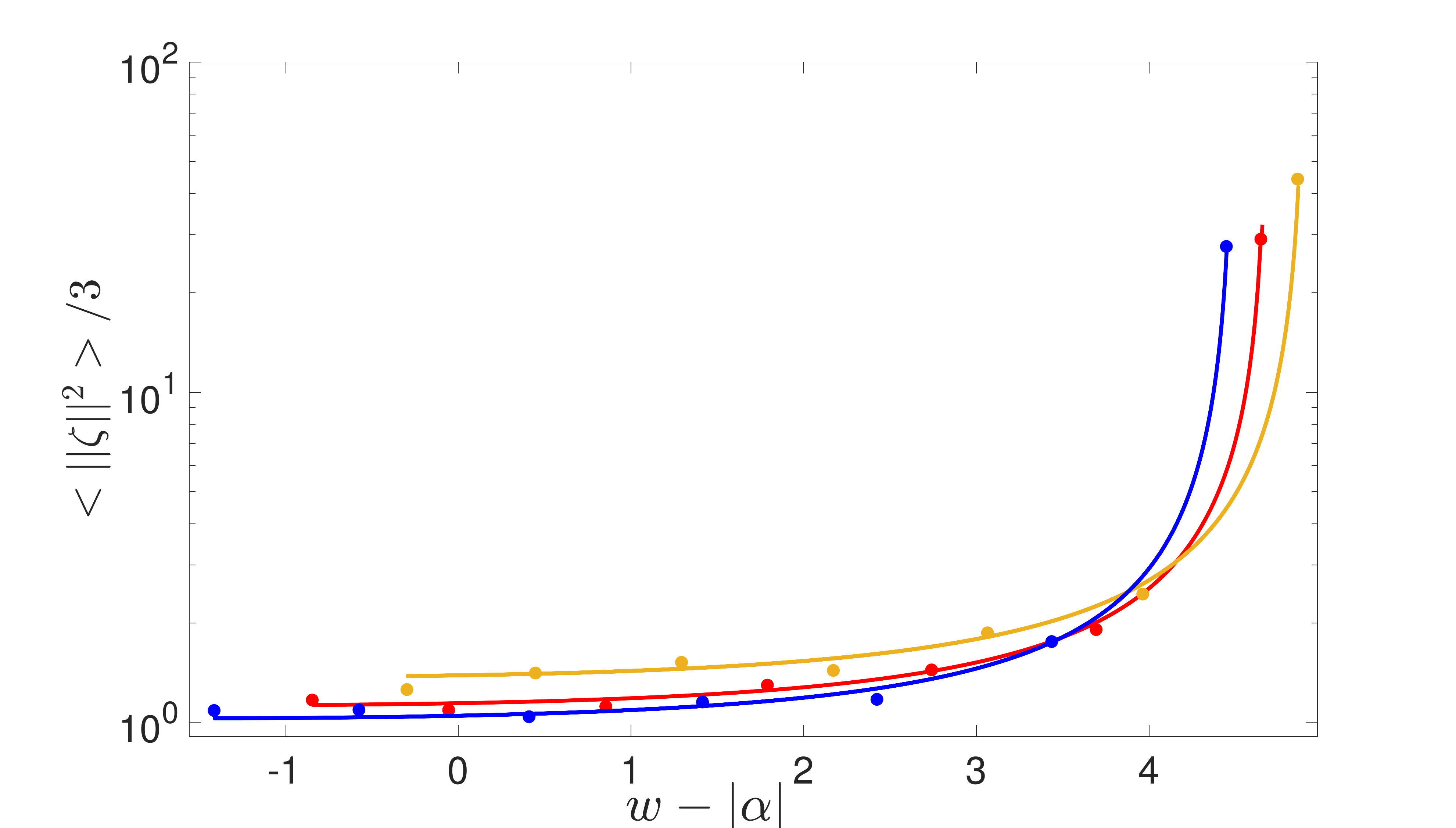}} \\
   \textbf{b}\\
  {\includegraphics[width=8.2cm]{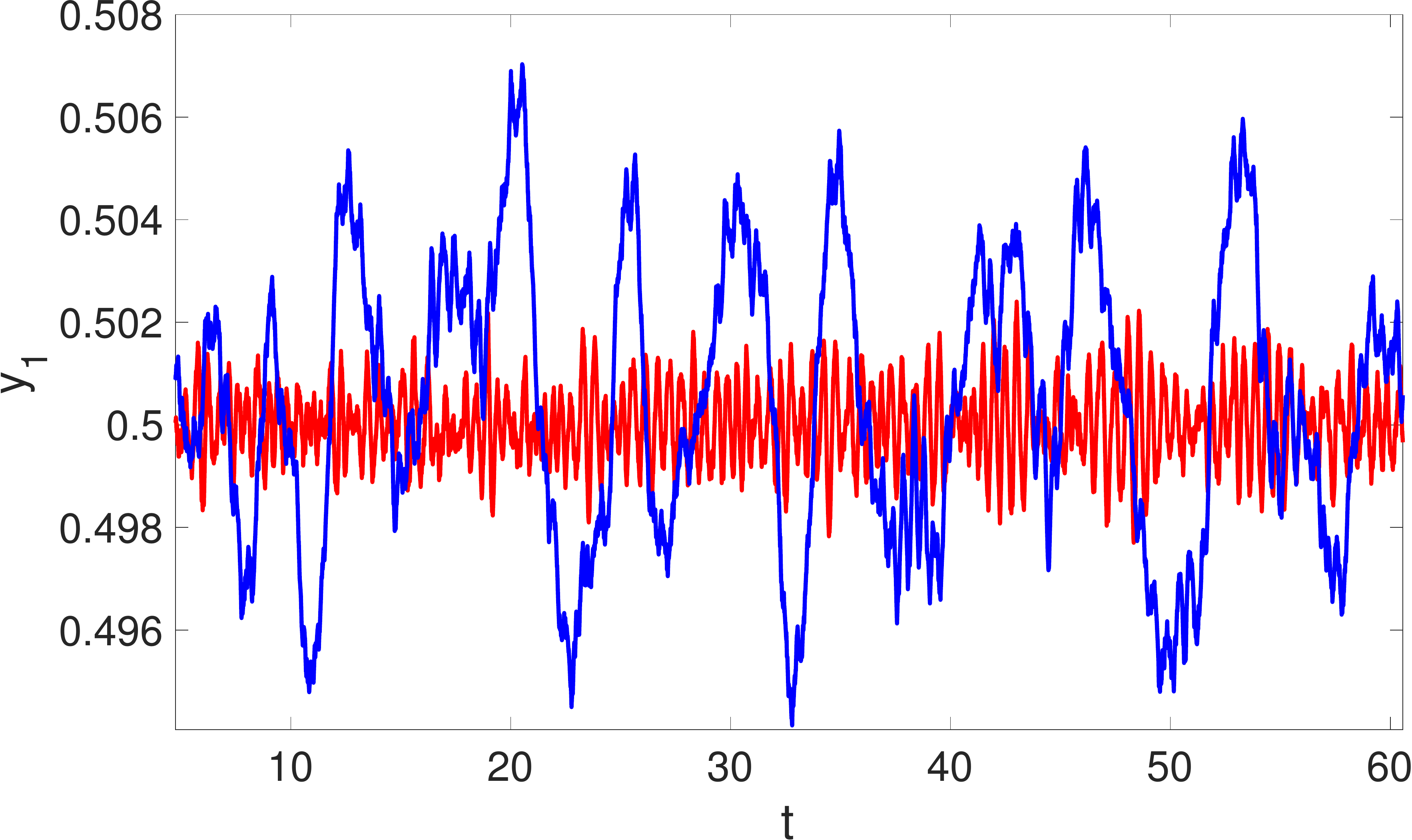}}
\end{tabular}

   \caption{Panel (a): the asymptotic norm of the fluctuations $<||\boldsymbol{\zeta}||^2>/3=\sum_{i=1}^3 (\xi_i^2+\eta_i^2)/3$, as displayed on each individual node, is plotted versus $w-|\alpha|$, moving along iso-lines $\bar{\alpha}$. Different curves refer to different choices of $\bar{\alpha}$ ($=-0.8,-0.6,-0.4$, from bottom to top). Solid lines stands for 
the analytical solution after equations (\ref{moments}). Symbols are obtained from direct simulations of the non linear Langevin equations 
(\ref{eq:eq Langevin non lineare vari nodi X}), averaging over $M=300$ independent realizations. Here, $V_1=V_2=V_3=10^6$ and $r=50$: Panel (b):  stochastic trajectories are displayed, relative to the inhibitors on the first node, i.e. species $y_1$, for different values of the  numerical abscissa. The red (small amplitude) trajectory is obtained for $w=-0.24$ ($\epsilon=1$), while the blue (large amplitude) trajectory refers to $w=5.18$ ($\epsilon=0.506$). Here $r=50$ and $V_1=V_2=V_3=10^6$.}
   \label{fig3}
  \end{figure}

In Figure \ref{fig3}, $ <||\boldsymbol{\zeta}||^2>/3=\sum_{i=1}^3 (\xi_i^2+\eta_i^2)/3$, the stationary norm of fluctuations on one node of the collection, is plotted against the reactivity index $w$, moving on (different) iso-$\alpha$ lines. Solid lines stand for the analytical calculations, as follows equations (\ref{moments}), while the symbols refer to the homologous quantities computed from direct simulation of the non linear Langevin equations (\ref{eq:eq Langevin non lineare vari nodi X}), via the Euler-Maruyama algorithm \cite{euler_maruyama}. 
The satisfying agreement between theory and simulations testify on the adequacy of the linear noise approximation. The positive correlation between $<||\boldsymbol{\zeta}||^2>/3$ and $w$, suggests that non normality controls the amplitude of emerging quasi cycles. The effect becomes more pronounced when $w>0$, i.e. when the reactivity of the non normal Jacobian  drives a self-consistent growth for the norm of the injected stochastic perturbation. Notice that $w$ is found to increase when crawling on the iso-$\alpha$ curves, from right to left,  in the plane ($\epsilon, D$): it is remarkable that the progressive gain in reactivity is triggered by a steady reduction in $\epsilon$, which implies forcing the system to be symmetric, at odd with intuition. Despite the fact that we have here chosen to display the cumulative contribution, the norm of both the activators and inhibitors species is found to grow, with the reactivity index $w$, when $w>0$. Moreover, the ensuing amplification can be made more conspicuous by differentiating the volumes $V_i$, across the loop. 

\begin{figure}[ht!]
\centering
{\includegraphics[width=9.5cm]{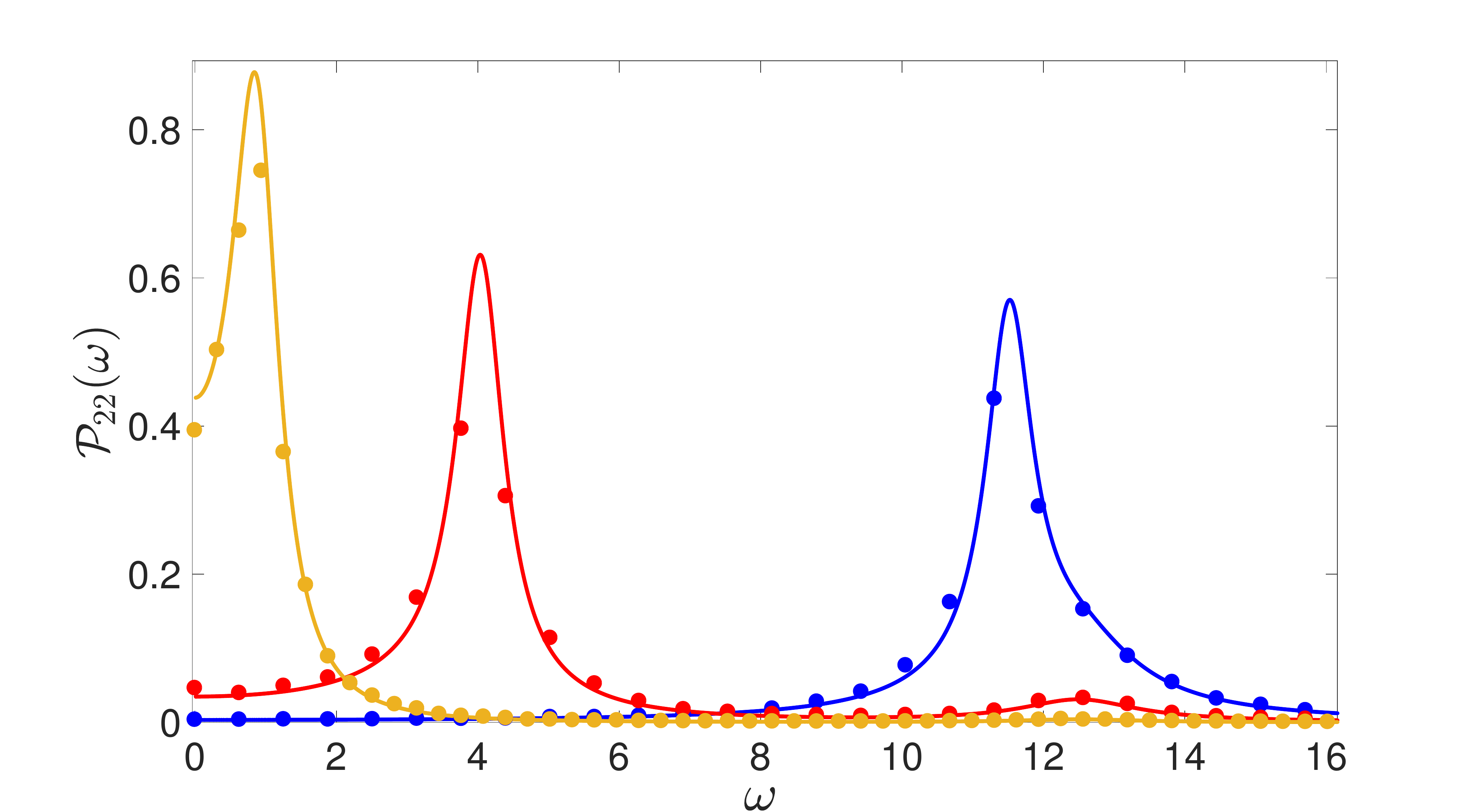}} \\
\caption{The theoretical power spectrum of the inhibitory species, $\mathcal{P}_{22}$, is plotted with a solid line against $\omega$. Different curves refer to different choices of ($D$,$\epsilon$) constrained to move across the iso-$\alpha$ line $\bar{\alpha}=-0.4$. The degree of reactivity, as quantified by the numerical abscissa $w$, increases from right to left ($w=0.1,4.6,5.2$): the peak of the power spectrum gains correspondingly in power. Symbols refer to direct numerical simulations, based on equations (\ref{eq:eq Langevin non lineare vari nodi X}), averaging over $M=200$ independent realizations. Here, $V_1=V_2=V_3=10^6$ and $r=50$. Notice that  $\mathcal{P}_{22}= \mathcal{P}_{44}=\mathcal{P}_{66}$, due to translational invariance across the loop. }
\label{ps} 
\end{figure}

\begin{figure}[ht!]
\centering
{\includegraphics[width=9cm]{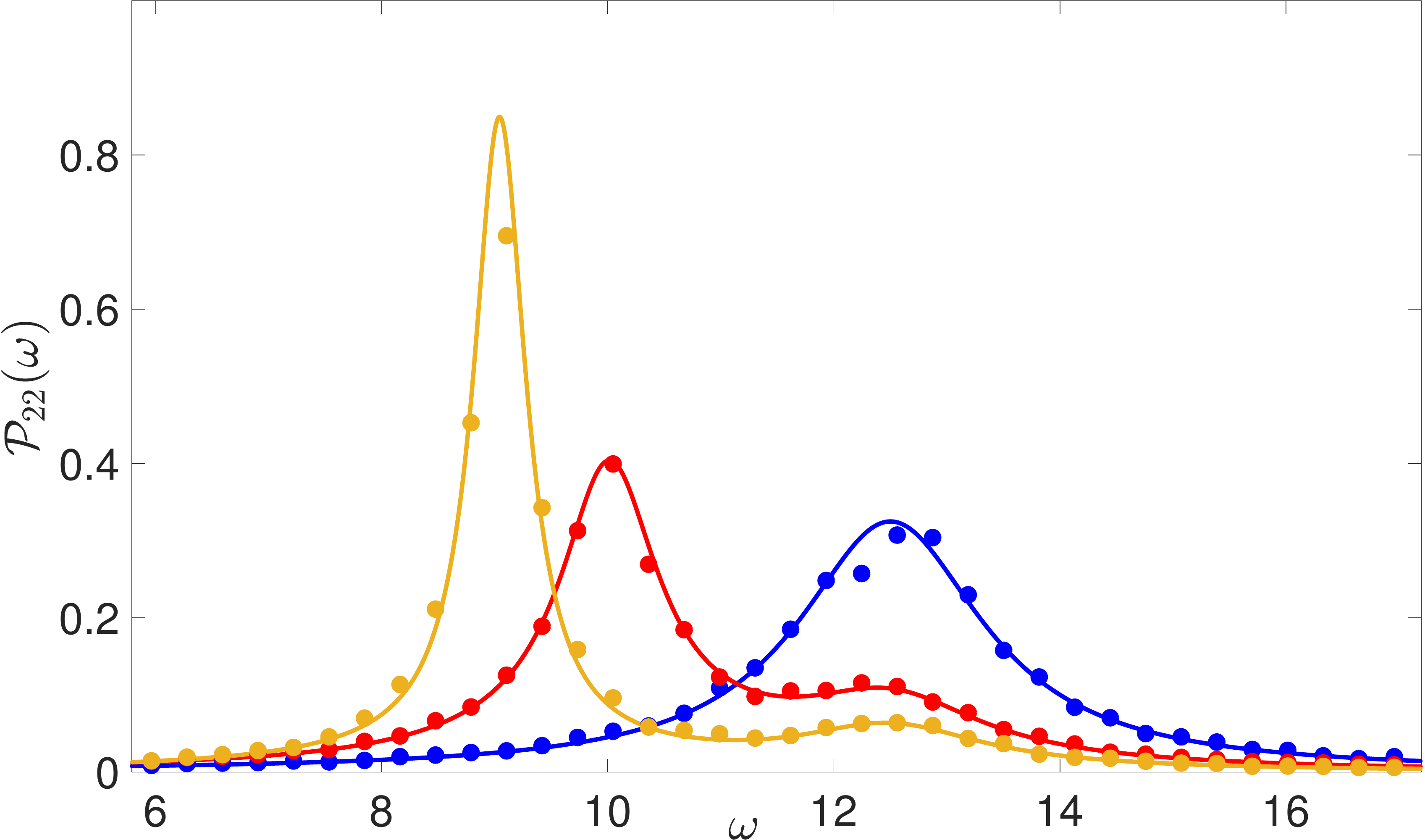}} \\
\caption{The theoretical power spectrum of the inhibitory species, $\mathcal{P}_{22}$, is plotted with a solid line against $\omega$, for different choices of $D$ (from right to left $D=0, 6, 8$). 
Symbols refer to direct numerical simulations, based on equations (\ref{eq:eq Langevin non lineare vari nodi X}), averaging over $M=200$ independent realizations. Here, $V_1=V_2=V_3=10^6$;  $\mathcal{P}_{22}= \mathcal{P}_{44}=\mathcal{P}_{66}$, due to translational invariance across the loop and $r=50$. }
\label{psD} 
\end{figure}

To further characterize the amplification of the stochastic cycles, against $w$, at fixed $\alpha$, we compute the power spectrum of the fluctuations around the deterministic fixed point. To this end we apply the temporal Fourier transform on both sides of \eqref{eqn:langevin} and  obtain the following equation
\begin{equation}\label{eqn:fourier}
-i\omega\tilde{\zeta}_i(\omega)=(\mathcal{J}\tilde{\mathbf{\zeta}})_i+\tilde{\lambda_i},
\end{equation}
where $\tilde{\zeta}$ stands for the Fourier transform of ${\zeta}$. Then, define the matrix $\Phi_{ij}=-i\omega\delta_{ij}-\mathcal{J}_{ij}$. The solution of \eqref{eqn:fourier} can be written as
\begin{equation}
\mathbf{\tilde{\zeta}}=\Phi^{-1}\tilde{\boldsymbol{\lambda}}.
\end{equation}
The power spectrum density matrix (PSDM) is consequently defined by the elements
\begin{equation}
    \mathcal{P}_{ij}(\omega)=<\tilde{\zeta}_i(\omega)\tilde{\zeta}^*_j(\omega)>.
\end{equation}
A straightforward calculation yields
\begin{equation}
\mathcal{P}_{ij}(\omega)=\left(\Phi^{-1}(\omega)\mathcal{B}(\Phi^{-1})^\dagger(\omega)\right)_{ij},
\end{equation}
whose diagonal elements represent the power spectra of the signals. 
In Figure \ref{ps}, three different power spectra, relative to the inhibitory species, are represented for distinct choices of the reactivity parameter $w$. When $w$ is made to increase, while keeping  $\alpha$ fixed, the power spectrum shifts towards the left, as prescribed by the formula for $\omega_1$, which sets the position of the peak. In agreement with the above, the peak gains in potency when the degree of reactivity is augmented.  Moving along iso-$\alpha$ lines is essential to prevent spurious contributions that might set in when the system is pushed towards the edge of the Hopf bifurcation. A gain of the quasi-cycles amplitude, is in fact observed when $\epsilon$ is kept constant and $D$ modulated in the range from $0$ to $D_c(\epsilon)$, as demonstrated in Figure \ref{psD}.  Although interesting per se, this phenomenon is, to a large extent, dictated by the progressive reduction in the value of $\alpha$, which is enforced by making $D$ approach its critical value $D_c$. Disentangling this latter contribution from the contextual raise in reactivity is arduous, and this is ultimately the reason why we have chosen to foliate the relevant parameters space in curves characterized by a constant damping factor $\alpha$.  Similar conclusion holds when monitoring the power spectra of fluctuations relative to the excitatory species.  

The above analysis suggests that the conversion of a stochastic input into regular oscillations is more efficient, in terms 
of amplification gain, when the reactivity of the system gets more pronounced. This observation provides an alternative angle to interpret the mechanism of noise driven amplification, as originally discussed in the \cite{zagli}. It can be in fact proven, that the Jacobian matrix that rules the self-consistent amplification as displayed in \cite{zagli} is non normal: its inherent reactivity grows with the coupling strength among adjacent nodes, i.e. with the parameter that boosts the exponential magnification of fluctuations along the unidirectional chain. In the setting explored in \cite{zagli},  the analogue of the damping factor $\alpha$ is always  constant and, as such, independent on the strength of the imposed coupling. This is at variance with the current implementation, where excitatory and inhibitory species are arranged on a triangular loop and iso-$\alpha$ curves are non linear functions of the parameters of the model. The intertwingled activity of excitatory-inhibitory populations gets self-consistently amplified by circulating the signal across a symmetric or asymmetric cyclic loop,  a minimal computational unit which constitutes the fundamental building block of any large networks, notwithstanding their diverse and variegated topology.  In the following, we will continue elaborating along this line and show, from a thermodynamical perspective, that the reactivity promotes the out-of-equilibrium dynamics of the scrutinized system. As such, it holds promise to result in an additional ingredient to lay the foundation of stochastic thermodynamics from the micro to the macro realms.

\section{Thermodynamics of a reactive loop.}

The goal of this section is to analyze the process of noise driven amplification across the circular loop from a thermodynamic point of view. In doing so we shall provide a novel angle to contextualize the implications of reactive non normality. To this end we recall that $\Pi(\boldsymbol{\zeta},\tau)$, the distribution of fluctuations  obeys to the Fokker-Planck equation (\ref{eqn:fokker}), in the linear noise approximation. Label with $f_i \equiv (\mathcal{J} \boldsymbol{\zeta})_i$ the non conservative forces that define the drift term in the aforementioned Fokker-Planck equation; $\mathcal{B}_i$ stands instead for the diffusive contribution. 

The Fokker-Planck equation (\ref{eqn:fokker}) can be written in the form of a continuity equation:
\begin{equation}
\frac{\partial \Pi}{\partial \tau}=-\boldsymbol{\nabla}\cdot \boldsymbol{\mathcal{I}}=-\sum_i \frac{\partial}{\partial \zeta_i}\mathcal{I}_i
\end{equation}
where we have defined the probability density current 
\begin{equation}
\mathcal{I}_i=f_i \Pi-\frac{\mathcal{B}_i}{2}\frac{\partial}{\partial 	\zeta_i}\Pi
\end{equation}
and use has been made of the fact that $B_i$ are constants.
In the limit $\tau \rightarrow \infty$, stationarity is achieved 
\begin{equation}
\frac{\partial \Pi}{\partial \tau}=0
\end{equation}
and the probability current is a solenoidal vector field
\begin{equation}
    \boldsymbol{\nabla}\cdot \boldsymbol{\mathcal{I}}=0.
\end{equation}
Equilibrium represents a very specific stationary solution, attained by imposing a vanishing probability current, namely 
$\boldsymbol{\mathcal{I}}\equiv 0$. Hence, 
\begin{equation}
    f_i=\frac{\mathcal{B}_i}{2}\frac{\partial}{\partial \zeta_i}\ln(\Pi).
\end{equation}
If we suppose that the system is in contact with just one thermal bath ($\mathcal{B}_i=\mathcal{B} \!\!\! \quad\forall i$), or restating the assumption in the context of interest, assuming that $\gamma_i=\gamma_1=1$ $\forall i$, the following consistency requirement should be matched: 

\begin{equation}
    \frac{\partial}{\partial \zeta_j}f_i=\frac{\partial}{\partial \zeta_i}f_j
\end{equation}
The above expression implies that the forces must be conservative, i.e. they can be obtained by a generalized potential $\mathcal{U}$ 
\begin{equation}
   f_i=-\frac{\partial}{\partial \zeta_i}\mathcal{U} 
\end{equation}
The definition of the current becomes therefore
\begin{equation}
    -\frac{\partial}{\partial \zeta_i}\mathcal{U}=\frac{\mathcal{B}}{2}\frac{\partial}{\partial \zeta_i}\ln(\Pi)
\end{equation}
and the above expression can be readily integrated to return the usual Boltzmann-Gibbs distribution
\begin{equation}
    \Pi(\boldsymbol{\zeta})=K\exp\left(-\frac{2}{\mathcal{B}}\mathcal{U}\left(\boldsymbol{\zeta}\right)\right)
\end{equation}
where $K$ stands for a proper normalisation constant. More interesting is the setting where the forces are non conservatives and the system evolves towards a stationary state, different from the conventional equilibrium. To explore this possibility we set to introduce the entropy functional $\mathcal{S}(\tau)$ from the probability distribution $\Pi(\boldsymbol{\zeta},\tau)$ as \cite{thermo0,thermo,thermo1}:

\begin{equation}
\label{entropy_def}
    \mathcal{S}(\tau)=-\int_V d\boldsymbol{\zeta}\Pi(\boldsymbol{\zeta},\tau)\ln\left(\Pi\left(\boldsymbol{\zeta},\tau \right)\right)
\end{equation}
where $V$ is the sample space of the dynamical variables. The Fokker-Planck equation sets the temporal evolution of the entropy. Taking the derivative of 
(\ref{entropy_def}) with respect to time $\tau$, and making use of the Fokker-Planck equation, one obtains \cite{thermo,thermo1}:
\begin{equation}\label{eq: prima eq derivata entropia}
    \frac{d\mathcal{S}}{d\tau}=-\int_V \frac{\partial \Pi}{\partial\tau}\left(\ln \Pi+1\right) d\boldsymbol{\zeta}=\int_V \sum_i \frac{\partial \mathcal{I}_i}{\partial \zeta_i} \left(\ln \Pi+1\right) d\boldsymbol{\zeta}.
\end{equation}
Assuming that the probability current vanishes at the boundary of the volume $V$, a simple integration by parts returns:
\begin{equation}
\frac{d\mathcal{S}}{d\tau}=-\sum_i  \int_V \mathcal{I}_i\frac{\partial}{\partial \zeta_i} \ln \Pi d\boldsymbol{\zeta}.
\end{equation} 
By recalling the definition of the probability current, one can write:
\begin{equation}
    \frac{\partial}{\partial \zeta_i}\ln \Pi=\frac{2}{\mathcal{B}_i}f_i-\frac{2}{\mathcal{B}_i}\frac{\mathcal{I}_i}{\Pi}.
\end{equation}
Finally, by substituting the above expression in the formula for the temporal evolution of the entropy, one eventually obtains \cite{thermo,thermo1}:
\begin{equation}
    \frac{d\mathcal{S}}{d\tau}=\varPi_S-\varPhi_S
\end{equation}
where 
\begin{equation}
    \varPi_S=\sum_i \frac{2}{\mathcal{B}_i}\int_V \frac{\mathcal{I}_i^2(\boldsymbol{\zeta},\tau)}{\Pi(\boldsymbol{\zeta},\tau)}d\boldsymbol{\zeta}
\end{equation}
and 
\begin{equation}
\varPhi_S=\sum_i \frac{2}{\mathcal{B}_i}\int_V f_i(\boldsymbol{\zeta})\mathcal{I}_i(\boldsymbol{\zeta},\tau) d\boldsymbol{\zeta}
\end{equation}
The quantity $\varPi_S$ is always positive and can be interpreted as the entropy production rate given by the non conservative forces $f_i$.
On the other hand, $\varPhi_S$ can be either positive or negative, and can be identified as the entropy flux rate. If $\varPhi_S>0$ the flux is from the system towards the environment, the opposite scenario corresponding to $\varPhi_S<0$.   By invoking the definition of the current and performing a few integrations by parts,
one derives a compact formula for the entropy flux rate:
\begin{widetext}
\begin{equation}\label{eq: flusso entropia computazionale}
    \begin{split}
    \varPhi_S &=\sum_i\frac{2}{\mathcal{B}_i}\int_V f_i \mathcal{I}_i d\boldsymbol{\zeta}=\sum_i\frac{2}{\mathcal{B}_i}\int_V \left( f_i^2 \Pi-\frac{\mathcal{B}_i}{2}f_i\frac{\partial}{\partial \zeta_i}\Pi\right)d\boldsymbol{\zeta}=\\
    &= \sum_i\frac{2}{\mathcal{B}_i}\int_V \left( f_i^2 \Pi+\frac{\mathcal{B}_i}{2}\Pi\frac{\partial}{\partial \zeta_i}f_i\right)d\boldsymbol{\zeta}= \sum_i \left( \frac{2}{\mathcal{B}_i} <f_i^2 >+<\frac{\partial}{\partial \zeta_i}f_i>\right)
    \end{split}
\end{equation}
\end{widetext}

which, by expliciting the forces $f_i$, yields:

\begin{widetext}
\begin{equation}
    \begin{split}
    \varPhi_S&= \sum_{i,j,k}\frac{2}{\mathcal{B}_i} \mathcal{J}_{ij}\mathcal{J}_{ik}\mathcal{C}_{jk}+\sum_{i,j,k}\frac{2}{\mathcal{B}_i} \mathcal{J}_{ij}\mathcal{J}_{ik}<\zeta_j><\zeta_k>+\sum_i\mathcal{J}_{ii}=\\
    &=\sum_i \frac{2}{\mathcal{B}_i}\left( \mathcal{J}\mathcal{C}\mathcal{J}^t\right)_{ii}+\sum_i \frac{2}{\mathcal{B}_i}\big[ \left(\mathcal{J}<\boldsymbol{\zeta}>\right)_i\big]^2+Tr\mathcal{J}.
    \end{split}
\end{equation}
\end{widetext}

In the stationary state:
\begin{equation}
    (\varPhi_S)_{\infty}=\sum_i \frac{2}{\mathcal{B}_i}\left( \mathcal{J}\mathcal{C}_s\mathcal{J}^t\right)_{ii}-6
\end{equation}
where use has been made of the fact that the first moments of the distribution $\Pi$ vanish in the stationary states and that  
$Tr\mathcal{J}=-2(1+\frac{1}{\gamma_2}+\frac{1}{\gamma_3})=-6$. Here, $C_s$ stands for the stationary  correlation matrix.
Similarly, following an analogous pathway, one can prove that:

\begin{widetext}
\begin{equation}
\begin{split}
    \varPi_S&=\sum_i \frac{2}{\mathcal{B}_i}\left(\mathcal{J}\mathcal{C}\mathcal{J}^t \right)_{ii}+2Tr\mathcal{J}+\frac{1}{2}\sum_i\mathcal{B}_i \mathcal{C}_{ii}^{-1}+\\
    &+\sum_i \frac{2}{\mathcal{B}_i}\left[ \left(\mathcal{J}<\boldsymbol{\zeta}> \right)_i\right]^2+4\left(\mathcal{J}<\boldsymbol{\zeta}> \right)_i\left(\mathcal{C}^{-1}<\boldsymbol{\zeta}> \right)_i+2\mathcal{B}_i\left[ \left(\mathcal{C}^{-1}<\boldsymbol{\zeta}> \right)_i\right]^2
\end{split}
\end{equation}
\end{widetext}

In the stationary state, for the setting of interest where the nodes share the same volume ($\gamma_i=\gamma_1$ $\forall i$), the entropy production rate matches the expression:

\begin{equation}
    (\varPi_S)_{\infty}=2Tr\left(\mathcal{J}\mathcal{C}_s\mathcal{J} \right)+2Tr\mathcal{J}+\frac{1}{2}Tr\mathcal{C}_s^{-1}.
\end{equation}

A straightforward, although lengthy, calculation confirms that $(\varPi_S)_{\infty}=(\varPhi_S)_{\infty}$, i.e. the condition for stationarity should be obviouvsly met. The entropy can be calculated, at any time $\tau$, by inserting in the definition (\ref{entropy_def}) the general solution of the Fokker-Planck equation (\ref{eqn:fokker}). This is the multivariate Gaussian of equation (\ref{eq: probabilità lineare}). Carrying out the calculations returns \cite{thermo2} :

\begin{widetext}
\begin{equation}
\label{entro_integral}
    \mathcal{S}(\tau)=\frac{1}{2}\int \Pi\left(\boldsymbol{\zeta},\tau\right) \left[ \left(\boldsymbol{\zeta}-<\boldsymbol{\zeta}>\right)^T\mathcal{C}^{-1}\left(\boldsymbol{\zeta}-<\boldsymbol{\zeta}>\right) +  \ln{\big(\left(2\pi\right)^{2\Omega}\abs{\mathcal{C}}}\big)  \right]d\boldsymbol{\zeta}
\end{equation}
\end{widetext}
The second term in the above integral gives simply $\ln{\big(\left(2\pi\right)^{2\Omega}\abs{\mathcal{C}}}\big)$, because of the normalisation of the probability distribution
The first term can be calculated as follows. Observe that $\mathcal{C}$ is a symmetric positive definite matrix and its elements are real. It is hence possible to construct its Cholesky decomposition $\mathcal{C}=EE^T$. Perform now the transformation $\boldsymbol{\zeta}=E\mathbf{s}+<\boldsymbol{\zeta}>$, which yields $\left(\boldsymbol{\zeta}-<\boldsymbol{\zeta}>\right)^T\mathcal{C}^{-1}\left(\boldsymbol{\zeta}-<\boldsymbol{\zeta}>\right)=\mathbf{s}^T\mathbf{s}$. The probability distribution expressed as a function of the variables $\mathbf{s}$ reads 
    \begin{equation}
        \Pi(\mathbf{s})=\frac{1}{\sqrt{(2\pi)^{2\Omega}}}\exp \left\{-\frac{1}{2}\mathbf{s}^T\mathbf{s} \right\}
    \end{equation}
    and consequently the first integral in (\ref{entro_integral})
 gives 
 
    \begin{equation}
         <\mathbf{s}^T\mathbf{s}>=2\Omega
    \end{equation}
 
In conclusion, $\mathcal{S}(\tau)=\frac{1}{2}\big[ 2\Omega+\ln\big( \left(2\pi\right)^{2\Omega}\abs{\mathcal{C}}  \big)  \big]=\frac{1}{2}\ln{\big(\left(2\pi e \right)^{2\Omega} \abs{\mathcal{C}}\big)}$ and 
$\mathcal{S}_{\infty}=\frac{1}{2}\ln{\big(\left(2\pi e \right)^{2\Omega} \abs{\mathcal{C}_s}\big)}$ with an obvious meaning of the symbols involved. 

\begin{figure}[ht!]
\centering
{\includegraphics[width=8.5cm]{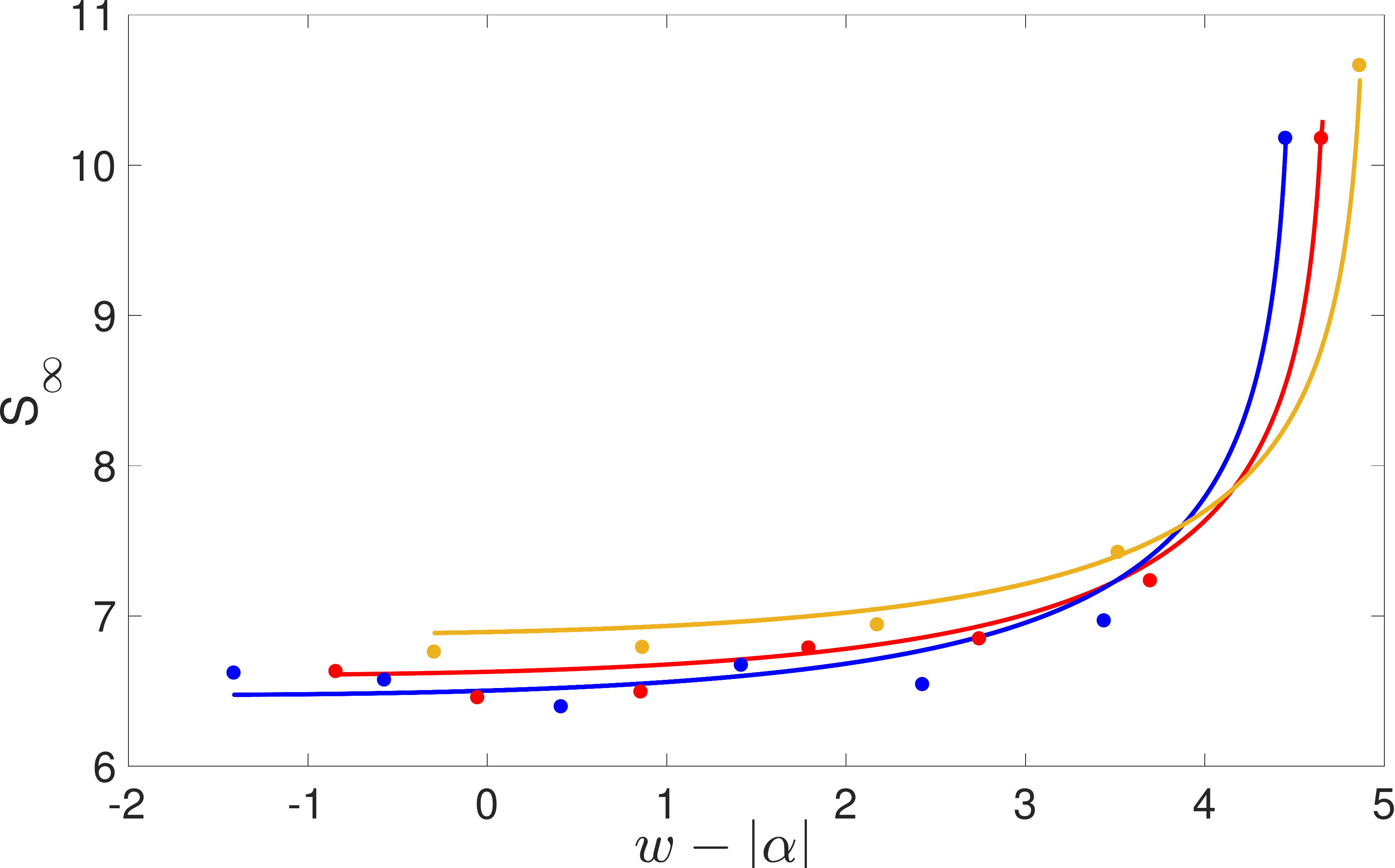}} \\
\caption{The stationary entropy $\mathcal{S}_{\infty}$ is plotted against $w$ moving on iso-$\alpha$ lines. Here, $\bar{\alpha}$ ($=-0.8,-0.6,-0.4$, from bottom to top). Solid lines stand for the analytical solutions.  Symbols follow from direct simulations of the Langevin equations 
(\ref{eq:eq Langevin non lineare vari nodi X}), upon averaging over $M=300$ independent realizations. Here, $V_1=V_2=V_3=10^6$ and $r=50$.}
\label{entropia1} 
\end{figure}

\begin{figure}[ht!]
\centering
{\includegraphics[width=8.8cm]{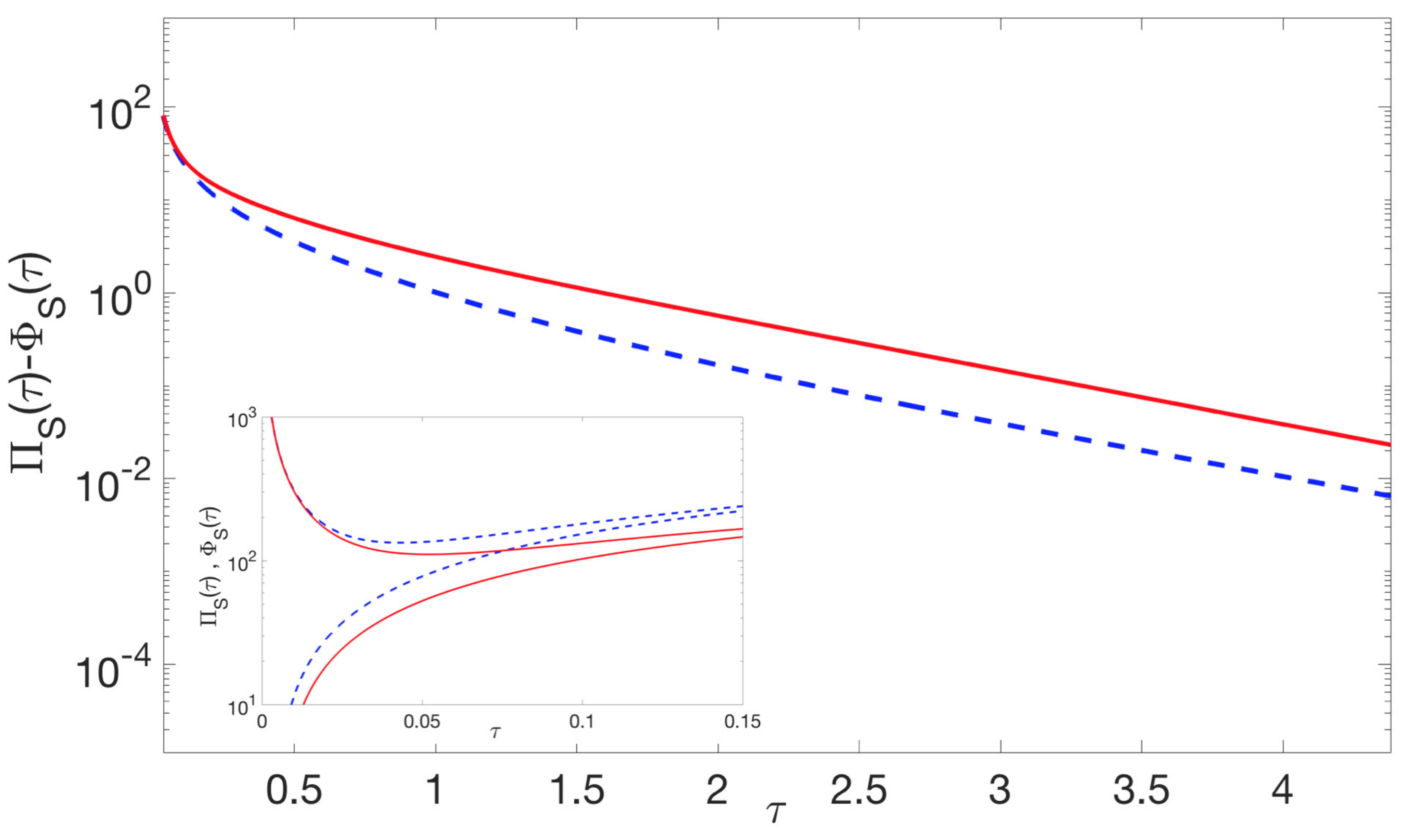}} \\
\caption{Main panel: $\varPi_S - \varPhi_S$ is plotted in lin-log scale against $\tau$, for two choices of the parameters $D, \epsilon$ constrained to return constant $\alpha=\bar{\alpha}=-0.6$.
The solid (red) line refers to $w=5.25$, while the dashed (blue) line stands for $w=-0.24$. In the inset $\varPi_S$ and $\varPhi_S$ are reported for the choices of parameters, as operated in the main panel.  Here, $r=50$.}
\label{entropia2} 
\end{figure}

In Figure \ref{entropia1} the stationary entropy $\mathcal{S}_{\infty}$ is plotted against $w$ moving on iso-$\alpha$ lines: the stationary entropy grows with the reactivity of the system. The reactivity, as stemming from non normality, facilitates hence the exploration of the available phase space, pushing the system out of equilibrium. In the transient phase, $\varPi_S> \varPhi_S$, as it can be appreciated in the main panel of Figure \ref{entropia2}, where $\varPi_S - \varPhi_S$ is represented against $\tau$. The two curves refer to different pairs $(\epsilon,D)$, chosen on the iso-$\alpha$ line $\bar{\alpha}=-0.6$. During the initial violent relaxation, the curves are almost indistinguishable, but then separate to  proceed on distinct tracks. More importantly, the out-of-equilibrium transient regime seems to persist for longer times, when the value of $w$ is made larger (solid vs. dashed lines). Indeed, the smaller $w$, the sooner the stationary condition $\varPi_S = \varPhi_S$ is established, as illustrated in the inset of Figure \ref{entropia2}. Here, the quantities $\varPi_S$  and $\varPhi_S$ are monitored as a function of time, in lin-log scale, for the same choice of parameters as in the main panel.

\section{Conclusions}

Finite size corrections represent an unavoidable source of endogenous disturbance, which can significantly impact the dynamics of the system under examination. Macroscopic order can materialize from the microscopic disorder, as stemming from the inherent demographic noise. Under specific operating conditions, quasi-cycles can develop via a resonant mechanism, triggered by the stochastic component of the dynamics. In general, quasi-cycles are modest in size and it is interesting to elaborate on the possible strategies, of either artificial or natural inspirations, that yield a coherent amplification of the stochastic signal. In a recent paper \cite{zagli}, it was shown that giant, noise assisted oscillations can develop when replicating a minimal model of excitatory and inhibitory units, on a large one dimensional lattice subject to unidirectional couplings. The parameters are assigned in such a way  that the deterministic analogue of the scrutinized stochastic model displays a stable homogenous equilibrium. Fluctuations generated by the microscopic granularity yield seemingly regular oscillations, with tunable frequency, which gain amplitude across the lattice. The rate of amplification is controlled by the coupling constant, among adjacent patches.  Motivated by this analysis, we have here considered a variant of the model discussed in \cite{zagli} to shed light onto the fundamental ingredients which cooperate for the onset of the amplification. The species are assigned to populate a spatially extended loop made of three nodes. Triangular loops define the simplest non trivial closed paths in large network complexes:  for this reason, it is instructive to elaborate on their putative role in assisting the stochastic amplification of quasi-cycles.  A sensible increase in the stochastic oscillations is indeed obtained when propagating the signal across the loop, while forcing the system in a region where the deterministic homogeneous fixed point proves stable. The larger the coupling constant the more pronounced the measured gain. When the coupling is made stronger, one approaches the boundary of stability for the underlying equilibrium: the damping of fluctuations is consequently reduced and this explains the increase of oscillations' amplitude against $D$. More interesting is the amplification detected when freezing the dispersion relation, i.e. setting to a constant the  largest (negative real part of the) eigenvalue of the Jacobian. In this case, the degree of amplification is controlled by the reactivity index,  a parameter that quantifies the short time growth of the norm of an imposed perturbation. The larger the reactivity of the non normal Jacobian matrix -- associated to the spatially extended system -- the more pronounced the stochastic driven oscillations. Non conservative forces push the system out of equilibrium and the stationary value of the entropy is found to increase with the reactivity, here measured by the numerical abscissa. Based on these observations, we argue that non normality, and, more specifically, reactivity,  should be thoroughly considered, when bridging stochastic dynamics and out-of-equilibrium thermodynamics. More than that, we want to remark that we are facing an important and unconventional thermodynamic scenario. In fact, in the presence of non conservative forces the system converges asymptotically to a genuine nonequilibrium steady state, after a transient during which the entropy production rate monotonically vanishes and the system reaches a maximum of the entropy. This shows that a variational principle based on entropy maximization is compatible with the presence of a nonzero (entropy) current. This is because in our model all nodes are subjected to the same effective temperature (i.e. $\gamma_i = \gamma_1=1$ $\forall i$). We conjecture that when assuming different values of the volumes of the nodes and of the corresponding temperatures, the system converges to the more standard scenario of another genuine nonequilibrium state, which is a consequence of the variational principle of minimization of the entropy production rate, constant at any node of the system. In conclusion, we have here shown that minimalistic loops of intertangled excitatory and inhibitory units might trigger a coherent amplification of the stochastic oscillations, as exhibited on each isolated patch. Moreover,  deterministic non normality should be maximized for the the stochastic system to grow giant coherent oscillations.

\section*{Acknowledgments}
The authors acknowledge financial support from H2020-MSCA-ITN-2015 project COSMOS  642563.

\end{document}